\documentclass[12pt,preprint]{aastex}

\def\bra{\langle}
\def\ket{\rangle}
\newcommand{\mc}{\mathcal}

\newcommand{\ov}{\overrightarrow}
\newcommand{\bl}{\textbf}
\def\myref{\bibitem{dummy}\vspace{-2pt}}

\begin{document}

\title{Faraday Rotation of the Cosmic Microwave Background \\ Polarization and
Primordial Magnetic Field Properties}

\author{L. Campanelli\altaffilmark{1}, A. D.
Dolgov\altaffilmark{2}, M. Giannotti\altaffilmark{3}, and F. L.
Villante\altaffilmark{4}}

\affil{INFN - Sezione di Ferrara, I-44100 Ferrara, Italy}
\affil{Dipartimento di Fisica, Universit\`a di Ferrara, I-44100
Ferrara, Italy}

\altaffiltext{1}{campanelli@fe.infn.it}
\altaffiltext{2}{ICTP, Strada Costiera 11, 31014 Trieste, Italy;
ITEP, Bol. Cheremushkinskaya 25, 113259 Moscow, Russia;
dolgov@fe.infn.it}
\altaffiltext{3}{giannotti@fe.infn.it}
\altaffiltext{4}{villante@fe.infn.it}

\begin{abstract}
Measurements of the Faraday rotation of the cosmic microwave
background radiation (CMBR) polarization could provide evidence
for the existence of primordial magnetic fields. The Faraday
rotation could also allow the study of some properties of these
fields. In this paper, we calculate the angular dependence of the
Faraday rotation correlator for different assumptions about the
spectral index and correlation length of the magnetic field. We
show that the helical part of the magnetic field does not make any
contribution to the correlator. We stress the importance of the
angular resolution of the detector in the Faraday rotation
measure, showing that it could severely reduce the effect, even
for a relatively large magnetic field correlation length.
\end{abstract}

\keywords{magnetic fields--cosmology:theory--cosmic microwave
background}

\section{Introduction}

Astronomical observations have revealed the presence of
large-scale magnetic fields in the Universe. They exist in all
types of galaxies (spiral, elliptical, barred, and irregular), in
galaxy clusters, and probably in superclusters, with correlation
lengths $\xi \sim$ Mpc and intensities $B \sim \mu$G. The presence
of magnetic fields in all gravitationally bound large-scale
structures could suggest that they have been created after
structure formation through astrophysical mechanisms (as, for
example, the ``Biermann battery''). On the other hand, the
detection of magnetic fields in galaxies at high redshifts, could
represent a strong hint that magnetic fields have been generated
in the early Universe ({\it i.e} before structure formation) by
microphysics processes (for a full discussion see Grasso \&
Rubinstein 2001; Widrow 2003; Dolgov 2001; Giovannini 2004).

If large-scale magnetic fields have a primordial origin, they
could have observable effects on the cosmic microwave background
radiation (CMBR). In particular, as discussed in (Kosowsky \& Loeb
1996), existence of a magnetic field at the last scattering
surface, corresponding to a present-day value of
$B_{0}\sim10^{-9}{\rm G}$, may induce measurable Faraday rotation
of the CMBR polarization. Larger magnetic fields, at the level
$B_{0}\sim10^{-8}{\rm G}$, may even depolarize the CMBR (Harari
{\it et al.} 1997), owing to differential Faraday rotation across
the last scattering surface.

Clearly the measure of a non-zero Faraday rotation in the CMBR
polarization would be extremely important, since it would support
the primordial origin of cosmic magnetic fields, and would provide
direct information about the physics of the early Universe.

In this paper, we analyze the Faraday rotation of the CMBR
polarization induced by a small ($B_{0}\sim10^{-9}{\rm G}$)
primordial stochastic magnetic field. In particular, we calculate
the angular dependence of the Faraday rotation measure (RM) maps
for different assumptions about the magnetic field spectral index
and correlation length. We consider the possibility of helical
primordial magnetic fields, showing that, contrary to the
suggestion of (Pogosian {\it et al.} 2002; Pogosian {\it et al.}
2003), the helical part of the magnetic field does not contribute
to the Faraday RM maps at any angle. Finally, we discuss the
importance of the detector angular resolution in the Faraday
rotation measure, showing that it could severely constrain the
actual possibility of observing the effect.

The plan of the paper is as follows. In Section 2, we briefly
review the properties of a stochastic homogeneous and isotropic
magnetic field and we define its characteristic properties. In
Section 3, we discuss the Faraday rotation effects and show that
Faraday RM maps do not depend on the helical part of the magnetic
field. In Section 4, we discuss the angular dependence of Faraday
RM maps for different assumptions on the magnetic field spectrum.
Finally, we summarize main results in Section 5.

\section{The magnetic field spectrum}

We consider primordial stochastic magnetic field
$\bl{B}(\eta,\bl{x})$, created before the matter-radiation
decoupling. Its Fourier transform, $\bl{B}(\eta,\bl{k})$, is
defined according to
\footnote{Here and in the following we use the natural system of
units in which $\hbar = c = 1$. We use bold characters to indicate
vectors, {\it i.e.} $\bl{a}\equiv\ov{a}$, while normal characters
are used to indicate vector moduli, {\it i.e.} $a = |\bl{a}|$.}
\begin{equation}
B_i(\eta,\bl{k}) = \int d^3\bl{x} \; e^{i \bl{k}\cdot\bl{x}}
\,B_{i}(\eta,\bl{x}) \, , \;\;\;\;\;\;\; B_i(\eta,\bl{x}) =
\frac{1}{(2\pi)^{3}}\int d^3\bl{k} \; e^{-i
\bl{k}\cdot\bl{x}}\,B_{i}(\eta,\bl{k}) \, .
\end{equation}
Here $\eta$ is the conformal time, $\bl{x}$ are comoving
coordinates, and $\bl{k}$ are the comoving wavenumbers.

Under assumption of flux conservation, the magnetic field scales
as $a^{-2}$, where $a$ is the cosmological scale factor (see, {\it
e.g.}, Grasso \& Rubinstein 2001). The magnetic field at arbitrary
time can thus be related to its present value by
$\bl{B}(\eta,\bl{x}) = \bl{B}_{0}(\bl{x})/a(\eta)^2$, where the
subscript $0$ indicates today's values and we normalize the scale
factor according to $a(\eta_{0}) = 1$.

We are interested in the effects of a statistically homogeneous
and isotropic magnetic field. This means that the correlation
tensor of the magnetic field, $C_{ij}(\bl{r}_1,\bl{r}_2) \equiv
\langle B_{i0}(\bl{r}_1) B_{j0}(\bl{r}_2) \rangle$, is a function
of $r=|\bl{r}_1-\bl{r}_2|$ only and, moreover, it transforms as an
$SO(3)$ tensor. In terms of the Fourier amplitudes of the field,
these conditions (together with the fact that the magnetic field
is a divergence-free field) translate into (see {\it e.g.} Monin
\& Yaglom 1975; Caprini {\it et al.} 2004)
\begin{equation}\label{corrK}
\langle B_{i0}(\bl{k}) B_{j0}(\bl{k}^\prime) \rangle =
\frac{(2\pi)^3}{2} \delta(\bl{k}+\bl{k}^{\prime})
[P_{ij}S(k)+i\varepsilon_{ijl}\hat{k}_l A(k)] ,
\end{equation}
where $ P_{ij}=\delta_{ij}-\hat{k}_i\hat{k}_j$,
$\varepsilon_{ijl}$ is the totally antisymmetric tensor and
$\hat{k}_{i}=k_{i}/k$. Here $S(k)$ denotes the symmetric part and
$A(k)$ denotes the antisymmetric part of the correlator. Usually
$S(k)$ is referred to as the {\it magnetic power spectrum}, being
related to the total magnetic energy $E_B=\frac{1}{2}\int
d^3x\,\bl{B}_0^2(\bl{x})$ (the volume is normalized to $V=1$) by
\begin{equation}\label{EBHB}
{\mc E}_B(k) = 2\pi k^2 S(k) \, , {\qquad E}_B = \int_0^\infty dk
\; {\mc E}_B(k) \, .
\end{equation}
On the other hand, $A(k)$ is referred to as the {\it helical power
spectrum}, being related to the magnetic helicity $H_B=\int
d^3x\,\bl{A}_{0}(\bl{x})\cdot\bl{B}_{0}(\bl{x})$ by (see {\it
e.g.} Vachaspati 2001):
\begin{equation}\label{HBHB}
{\mc H}_B(k) = 4\pi k A(k) \, , {\qquad H}_B = \int_0^\infty dk \;
{\mc H}_B(k) \, .
\end{equation}
We will assume that both $S(k)$ and $A(k)$ can be represented by
the following simple functions:
\begin{equation}
\label{spectrum} S(k) = S_0 k^{n_{S}} e^{-(k/K)^2}, {\qquad A}(k)
= A_0 k^{n_{A}} e^{-(k/K)^2} ,
\end{equation}
which for $k\ll K$ possess a power law behavior. For large $k$,
the spectrum is instead suppressed exponentially in order to have
finite energy and helicity and a finite correlation length $\xi$,
given by:
\begin{equation}
\label{xi} \xi \equiv \frac{\int_0^\infty dk \, (2\pi/k) \, {\mc
E}_B(k)} {\int_0^\infty dk \, {\mc E}_B(k)} =
\frac{2\pi}{K}\frac{\Gamma[(n_{S} + 2)/2]}{\Gamma[(n_{S}+3)/2]} \,
.
\end{equation}
The two functions $S(k)$ and $A(k)$ are not completely
independent, since any field configuration has to satisfy:
\begin{equation}\label{relAS}
S(k)\geq |A(k)| .
\end{equation}
Moreover, requiring that the function (\ref{corrK}) is analytic,
one can show that the spectral index $n_{S}$ has to be even $\geq
2$, while $n_{A}$ has to be odd and $\geq 2$ (Durrer \& Caprini
2003).

\section{Faraday Rotation}

As discussed in (Kosowsky \& Loeb 1996), a primordial magnetic
field could leave significant imprints upon the CMBR polarization
through the effect of the Faraday rotation. Any magnetic field
between the last scattering surface and the observer would rotate
the polarization vector by the angle
\footnote{Here and in the following, we use the Heaviside-Lorentz
electromagnetic units in which the fine structure constant is
$\alpha = e^2/4 \pi$. In the previous version of this paper and in
the version accepted for publication in ApJ, pre-factors in some
formulae are incorrect. However, the final results are correct. We
thank T. Kahniashvili for making us aware of this point.}:
\begin{equation}
\label{Faraday} d\Phi = \lambda^2 \frac{e^3}{8\pi^2 m_e^2} \; n_e
\, \bl{B} \cdot \hat{\bl n} \, a \, d\eta \, ,
\end{equation}
where $\hat{\bl n}$ indicates the photon propagation direction,
$a$ is the scale factor, $\lambda$ is the photon wavelength and
$n_{e}$ is the number density of free electrons. In the assumption
of flux conservation, the quantity $\lambda^{2}\bl{B}$ is time
independent and thus one can substitute
$\lambda\rightarrow\lambda_{0}$ and ${\bl B} \rightarrow {\bl
B_{0}}$ into equation~(\ref{Faraday}).

The Faraday rotation is proportional to the number density of free
electrons, which evolves because of the Universe expansion and
because of recombination and reionization phenomena. It is useful
to describe these variations in terms of their contribution to the
photon optical depth. Following (Kosowsky \& Loeb 1996), we
introduce the differential optical depth
$\dot{\tau}=d\tau/d\eta=n_e\sigma_T a$ where $\sigma_{T} = 8 \pi
\alpha^2 / \, 3 m_e^2$ is the Thomson cross section. In terms of
this quantity, one has
\begin{equation}\label{faraday1}
d\Phi = \frac{3}{4 \pi e} \, \dot{\tau}(\eta) \,\lambda^2 _{0}\,
\bl{B}_{0} \cdot \hat{\bl{n}}\,d\eta \, .
\end{equation}
This expression has to be integrated along the photon path,
starting from the photon last scatter $\eta$ until the present
time $\eta_{0}$. One obtains
\begin{equation}\label{faraday2}
\Delta\Phi(\eta) = \lambda_0^2 \, \varrho(\eta) \, ,
\end{equation}
where $\varrho(\eta)$ is given by
\begin{equation}
\varrho(\eta) = \frac{3}{4 \pi e} \int^{\eta_{0}}_{\eta} \! d\eta'
\; \dot{\tau}(\eta') \, \bl{B}_{0}({\hat{\bl
n}}(\eta'-\eta_{0}))\cdot \hat{{\bl n}} \, .
\end{equation}
In the previous expression we have assumed that the observer is at
the origin of our reference frame and ${\hat{\bl
n}}(\eta'-\eta_{0})$ describes the photon space trajectory as a
function of the conformal time $\eta$.

As a final step, one takes into account the fact that photons from
a given direction last scattered at different times $\eta$. One
introduces, then, the visibility function
$g(\eta)=\dot{\tau}(\eta)\exp(-\tau(\eta))$, which gives the
probability that a photon observed at $\eta_{0}$ last scattered
within $d\eta$ of a given $\eta$, and calculates
\begin{equation}\label{RM0}
{\rm RM} = \int_{0}^{\eta_{0}} \! d\eta \, g(\eta) \,
\varrho(\eta) = \frac{3}{4 \pi e}\int_{0}^{\eta_{0}} \! d\eta \,
g(\eta) \int^{\eta_{0}}_{\eta} \! d\eta' \, \dot{\tau}(\eta') \,
\bl{B}_{0}({\hat{\bl n}}(\eta'-\eta_{0}))\cdot \hat{\bl{n}} \, .
\end{equation}
It is straightforward to recast Eq.~(\ref{RM0}) into the form
\begin{equation}\label{RM}
{\rm RM} = \frac{3}{4 \pi e}\int^{\eta_{0}}_{0} \! d\eta' g(\eta')
\, \bl{B}_{0}({\hat{\bl n}}(\eta'-\eta_{0}))\cdot \hat{\bl{n}} \,
,
\end{equation}
which gives the wavelength-independent Faraday rotation measure RM
in terms only of $g(\eta)$ and of the properties of the magnetic
field ${\bl B}_{0}$.

In principle, by measuring the CMBR polarization at different
wavelengths with a suitable accuracy, one should be able to
determine the Faraday RM as a function of the observation
direction and, thus, to determine the correlator
\begin{equation}
RR'(\theta) = \langle {\rm RM}({\hat{\bl n}}) {\rm RM}({\hat{\bl
m}}) \rangle \, ,
\end{equation}
where $\hat{\bl n}$ and $\hat{\bl m}$ are two directions on the
sky and $\cos\theta=\hat{\bl n}\cdot\hat{\bl m}$. This correlator
depends on the properties of the magnetic field and on the
ionization history of the Universe, according to
\begin{equation}
\label{RRformal} RR'(\theta) = \left(\frac{3}{4 \pi e}\right)^{2}
\int \! d\eta \, g(\eta) \int \! d\eta' g(\eta') \, \langle ({\bl
B}_{0} (\Delta \eta \;\hat{\bl n}) \cdot \hat{\bl n}) ({\bl B}_{0}
(\Delta \eta^\prime \;\hat{\bl m})\cdot \hat{\bl m}) \rangle \, ,
\end{equation}
where $\Delta \eta = \eta - \eta_{0}$ and $\Delta \eta^\prime =
\eta^\prime - \eta_{0}$. The last terms in the previous expression
can be expressed in Fourier space as
\begin{eqnarray}
\nonumber \langle (\bl{B}_{0}\cdot \hat{\bl n}) (\bl{B}_{0}\cdot
\hat{\bl m}) \rangle \!\!& = &\!\! \frac{1}{2(2\pi)^3} \int
d^{3}{\bl k} \; \left\{ \left[  ( \hat{\bl n} \cdot \hat{\bl m} )
      - (\hat{\bl n} \cdot \hat{\bl k}) (\hat{\bl m} \cdot \hat{\bl k}) \right]
 S(k) \right. \\
 \!\!& + &\!\! \left. i \left[(\hat{\bl n} \times \hat{\bl m})
\cdot \hat{\bl k}\right] A(k) \right\} \exp \left[ -i{\bl k} \cdot
\left( {\hat{\bl n}} \Delta \eta - {\hat{\bl m}} \Delta \eta'
\right) \right] . \label{B0B0}
\end{eqnarray}
We remind the reader that $S(k)$ and $A(k)$ are, respectively,
symmetric and helical terms of the magnetic field correlation
tensor.

In principle, in order to calculate rigorously the Faraday RM
correlator $RR'(\theta)$, one should solve the radiative transfer
equations for the microwave background polarization in the
presence of a magnetic field. The exact value of the rotation
measure is, in fact, sensitive to the growth history of the CMBR
polarization through the surface of the last scattering. This was
done, {\it e.g.}, in (Kosowsky \& Loeb 1996) where the quantity
$RR'(0)$ was calculated in the limit of small magnetic fields
({\it i.e.} $B_{0}\sim 10^{-9}$G), showing that, in this limit,
the simple approach described above gives reasonably accurate
results. In this paper, we use Eqs.~(\ref{RRformal},\ref{B0B0}),
which has the advantage of simplicity and allows us to calculate
analytically the angular dependence of $RR'$. This allows, in
turn, to discuss in a transparent way the dependence of the
Faraday RM maps on the magnetic field parameters, the effects of
finite detector angular resolution and a possible role of helicity
in $RR'$.

In this respect, it was suggested by Pogosian {\it et al.} (2002,
2003) that the dependence of $RR'$ on $A(k)$ can be used to
extract information on the helical part of the magnetic field.
However, one can show that the helical part of the magnetic field
does not contribute to $RR'$ at any angle $\theta$
\footnote{For $\theta = 0$ this follows simply from the fact that
helicity affects only off-diagonal terms of the magnetic field
autocorrelation function, as noted in (En\ss lin \& Vogt 2003).}.
This is due to the fact that its contribution to the magnetic
field correlator $\langle (\bl{B}_{0}\cdot \hat{\bl n})
(\bl{B}_{0}\cdot \hat{\bl m}) \rangle$, given by
\begin{equation}
\langle (\bl{B}_{0}\cdot \hat{\bl n}) (\bl{B}_{0}\cdot \hat{\bl
m})\rangle_{\rm hel} \equiv \frac{i}{2(2\pi)^3} \! \int d^{3}{\bl
k} \left( \hat{\bl n} \times \hat{\bl m} \cdot \hat{\bl k} \right)
A(k) \exp \! \left[ -i{\bl k} \cdot \left( {\hat{\bl n}} \Delta
\eta - {\hat{\bl m}} \Delta \eta' \right) \right] ,
\end{equation}
is zero for any chosen directions $\hat{\bl n}$ and $\hat{\bl m}$.
Indeed, by decomposing ${\bl k}$ in terms of ${\bl k}_{\perp}$ and
${\bl k}_{\parallel}$, where ${\bl k}_{\perp}$ and ${\bl
k}_{\parallel}$ are perpendicular and parallel, respectively, to
the plane containing the vectors $\hat{\bl n}$ and $\hat{\bl m}$,
we get
\begin{equation}
\langle (\bl{B}_{0}\cdot \hat{\bl n}) (\bl{B}_{0}\cdot \hat{\bl
m})\rangle_{\rm hel} = \frac{i}{2(2\pi)^3} \! \int \! d^3 {\bl k}
\left( \hat{\bl n} \times \hat{\bl m} \cdot \frac{\bl
k_{\perp}}{k} \right ) A(|{\bl k}_{\perp}+{\bl k}_{\parallel}|)
 \exp \! \left[ -i{\bl k}_{\parallel} \! \cdot \!
\left( {\hat{\bl n}} \Delta \eta - {\hat{\bl m}} \Delta \eta'
\right) \right] \! .
\end{equation}
One immediately sees that the integrand is odd in ${\bf
k}_{\perp}$, and thus the integral has to vanish. All this shows
that CMBR Faraday RM maps cannot give any information on the
helicity of primordial magnetic fields
\footnote{During the writing of this paper, we learned from
(Kahniashvili 2004) that Kosowsky {\it et al.} are also studying
this problem with similar conclusions.}.

In conclusion, Faraday RM maps of the CMBR only depends on the
symmetric part of the magnetic field and we have
\begin{equation}\label{RRprimo}
RR'(\theta) = \left(\frac{3}{4 \pi e}\right)^{2} \int
d\eta\;g(\eta) \int d\eta'\;g(\eta') \langle ({\bl B}_{0} (\Delta
\eta \hat{\bl n}) \cdot \hat{\bl n}) ({\bl B}_{0} (\Delta \eta'
\hat{\bl m})\cdot \hat{\bl m})\rangle_{\rm symm} \, .
\end{equation}
The magnetic field correlator can be expressed as (Kolatt 1997)
\begin{equation}\label{BBsymm}
\langle ({\bl B}_{0} \cdot \hat{\bl n}) ({\bl B}_{0} \cdot
\hat{\bl m}) \rangle_{\rm symm} = \left[  ( \hat{\bl n} \cdot
\hat{\bl m} ) C_{\perp}(r)
      + (\hat{\bl n} \cdot \frac{{\bl r}}{r}) (\hat{\bl m} \cdot \frac{{\bl r}}{r})
        (C_{\parallel}(r)-C_{\perp}(r)) \right] ,
\end{equation}
where ${\bl r}=\hat{{\bl n}} \Delta \eta - \hat{{\bl m}} \Delta
\eta'$, and
\begin{eqnarray}
\label{c-perp} C_{\perp}(r)  \!\!& = &\!\!
\frac{2}{3(2\pi)^3}\int_{0}^{\infty} \! dk \:
{\mc E}_{\rm B}(k) \left[ j_{0}(kr) - \frac12 j_{2}(kr) \right ] , \\
\label{c-parallel} C_{\parallel}(r) \!\!& = &\!\!
\frac{2}{3(2\pi)^3}\int_{0}^{\infty} \! dk \: {\mc E}_{\rm B}(k)
\left[ j_{0}(kr) +  j_{2}(kr) \right] ,
\end{eqnarray}
where $j_{i}(x)$ are the spherical Bessel functions of the
$i^{th}$ order.

\section{Results}

The bulk of the Faraday rotation is generated close to $\eta_{\rm
dec}$ where the photon visibility function $g(\eta)$ is maximal.
If the correlation length $\xi$ of the magnetic field is much
larger than the thickness of last scattering surface $\delta
\eta_{\rm dec}\sim 10\;{\rm Mpc}$ (Spergel {\it et al.} 2003)
({\it i.e.} the distance travelled by a photon during the period
of time in which $g(\eta)$ is sizeably different from zero), one
can approximate the visibility function with delta function
$g(\eta)=\delta(\eta-\eta_{\rm dec})$ and extract $\langle ({\bl
B}_{0}\cdot {\bl n}) ({\bl B}_{0}\cdot {\bl m})\rangle_{\rm symm}$
from the two integrals in Eq.~(\ref{RRprimo}). As a final result,
one obtains
\begin{equation}
\label{RRideal} RR'(\theta) = \left(\frac{3}{4 \pi e}\right)^{\!
2} \left \{ \hat{\bl n} \cdot \hat{\bl m} \, C_{\perp}(r_{\rm
dec}) + (\hat{\bl n} \cdot \frac{{\bl r}_{\rm dec}}{r_{\rm dec}})
(\hat{\bl m} \cdot \frac{{\bl r}_{\rm dec}}{r_{\rm dec}}) \,
[C_{\parallel}(r_{\rm dec})-C_{\perp}(r_{\rm dec})] \right \},
\end{equation}
where ${\hat {\bl n}}\cdot {\hat {\bl m}} =\cos \theta$ and ${\bl
r}_{\rm dec}=({\hat {\bl n}}-{\hat {\bl m}})(\eta_{\rm
dec}-\eta_{0})$.

For the case $\theta = 0$, expression~(\ref{RRideal}) can be
easily evaluated. We have
\begin{equation}
\label{RR0} RR'(0) = \frac{6}{(2\pi)^3 (4 \pi e)^{\, 2}} \int \!
dk \: {\mc E}_{\rm B}(k)= \frac{3}{(4 \pi e)^{\, 2}} \,
\overline{B}_{0}^{2} \, ,
\end{equation}
where $\overline{B}_{0}$ is the average magnetic field, defined by
\begin{equation}\label{B0A}
\overline{B}_{0}^2 = \int \! \frac{d^3\bl{k}}{(2\pi)^3} \,
\frac{d^3\bl{k}^\prime}{(2\pi)^3} \, \bra B_{0}(\bl{k})\cdot
B_{0}(\bl{k}^\prime)\ket \, = \, \frac{2}{(2\pi)^3} \, E_{B}.
\end{equation}
This result essentially coincides with the result of (Kosowsky \&
Loeb 1996) and corresponds to the average rotation of the CMBR
polarization approximately equal to
\begin{equation}
\label{Phi} \Phi = RR'(0)^{1/2} \lambda_{0}^2 \, \simeq \, 1.3^{o}
\left(\frac{B_{0}}{10^{-9}{\rm Gauss}}\right)
\left(\frac{\nu_{0}}{30 {\rm GHz}}\right)^{-2} \! .
\end{equation}

For $\theta\neq 0$, the situation is slightly more complicated.
After some algebra we obtain
\begin{equation}
\label{RRtheta} RR'(\theta) = \left(\frac{3}{4 \pi e}\right)^{\!
2} \left[ \, C_{\perp}(r_{\rm dec}) \cos^2 (\theta/2) \; - \;
C_{\parallel}(r_{\rm dec}) \sin^{2} (\theta/2) \, \right] ,
\end{equation}
where $r_{\rm dec} = 2(\eta_{0}-\eta_{\rm dec}) \sin
({\theta}/{2})$. However, in a realistic situation, the
correlation length $\xi$ of the magnetic field is much smaller
than the distance to the last scattering surface,
$\eta_{0}-\eta_{\rm dec}$. This permits us to consider the limit
of small observation angles ({\it i.e.} $\theta\ll 1$), in which
the second term in the right-hand side of Eq.~(\ref{RRtheta})
becomes negligible and one obtains
\begin{equation}
\label{RRdsmall} \frac{RR'(\theta)}{RR'(0)} =
\frac{C_{\perp}(\theta(\eta_{0}-\eta_{\rm dec}))} {C_{\perp}(0)}
\, .
\end{equation}
We remark that this expression is quite natural if we consider the
physical meaning of the function $C_{\perp}(r)$. It describes the
variation of the correlator $\langle B_{i}(0) B_{i}({\bl
r})\rangle=C_{\perp}(r)$ of the magnetic field components in a
fixed direction (which in our case is the photon propagation
direction), when ${\bl r}$ moves in a plane perpendicular to that
direction ({\it i.e.} on the last scattering surface).

To obtain an explicit expression for $RR'(\theta)$ one has now to
specify ${\mc E}_{\rm B}(k)=2\pi k^2 S(k)$ in Eq.~(\ref{c-perp}).
If $S(k)$ can be described by the simple functions given by
Eq.~(\ref{spectrum}), the integration can be performed
analytically. For the case $n_{\rm S}=2$ we get
\begin{equation}
\label{RRdelta} \frac{RR'(\theta)}{RR'(0)} = e^{-\chi^2\theta^2}
\left(1-\chi^2\theta^2\right) ,
\end{equation}
where $\chi=K (\eta_{0}-\eta_{\rm dec})/2$. For different $n_{S}$
values one obtains quite similar expressions, of the kind
$\exp(-\chi^{2}\theta^{2})P(\chi^{2}\theta^{2})$, where
$P(\chi^2\theta^{2})$ are higher order polynomials in the
$\chi^2\theta^2$ variable [in the Appendix we present explicit
expressions of $C_{\perp}(r)$ and $C_{\parallel}(r)$ for arbitrary
values of $n_{S}$]. Expression~(\ref{RRdelta}) can be explicitly
written as a function of the magnetic field correlation length
$\xi$, which for $n_{S}=2$ is $\xi = 8\sqrt{\pi}/3K$. For
different $n_{S}$ values, one has to take into account that the
relation between $\xi$ and $K$ depends on the chosen spectral
index, as it is described in Eq.~(\ref{xi}).

\begin{figure}[htb]
\label{fig1} \plotone{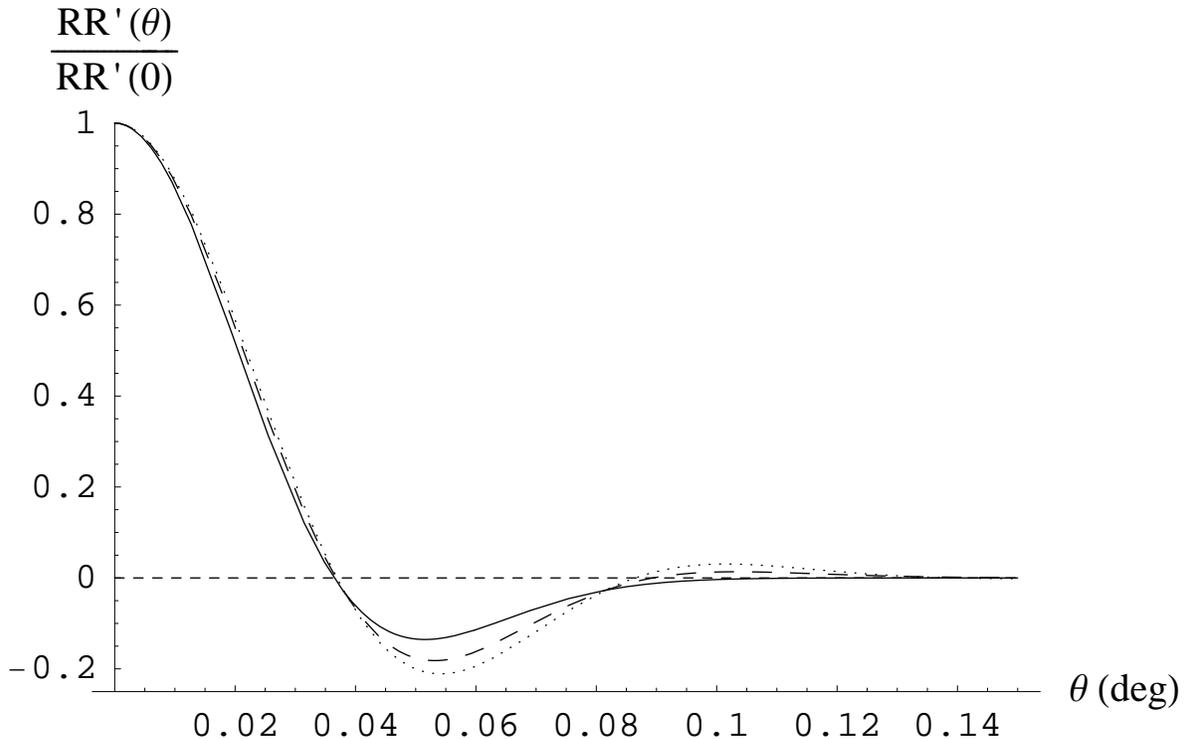} \caption{Faraday rotation measure
correlation $RR'(\theta)$ as a function of the separation angle
$\theta$. The three lines correspond to the magnetic field
spectral index $n_S =2$ ({\it solid line}), $n_S = 4$ ({\it dashed
line}) and $n_S = 6$ ({\it dotted line}). The correlation length
of the magnetic field is $\xi = 20$ Mpc.}
\end{figure}

In Fig.~1 we show $RR'$ as a function of the angle $\theta$
(normalized to its value at $\theta = 0$) for various choices of
the spectral index $n_{S}$ and for a fixed correlation length
$\xi=20$ Mpc. The three lines correspond to $n_{S}=2,4$, and $6$.
One can see that the function $RR'(\theta)$ has quite peculiar
behavior. In particular, in all cases, there are angles that
correspond to negative values of $RR'$. This reflects the behavior
of the magnetic field at the last scattering surface. Namely, it
is due to the fact that the correlator $\langle B_{\gamma}(0)
B_{\gamma}(r)\rangle \simeq C_{\perp}(r)$ (where $\gamma$
indicates the photons propagation direction) on the surface of the
last scatter becomes negative. In other words, the magnetic field
components $B_{\gamma}$ in different regions of the last
scattering surface can be anticorrelated.

From the point of view of observations, it is important to note
that the angular behavior of $RR'(\theta)$ is only marginally
dependent on the $n_{S}$ value. In particular, the point at which
$RR'(\theta)$ vanishes practically coincides in all three cases.
On one hand, this indicates that the observation of $RR'(\theta)$
could provide information on the correlation length $\xi$ that is
essentially independent of the parameter $n_{S}$. On the other
hand, it indicates that a large sensitivity is required to
discriminate among different $n_{S}$ values.

The results presented above are obtained neglecting the last
scattering surface thickness and assuming that the detector has a
perfect angular response. These approximations are correct if the
magnetic field correlation length $\xi$ is much larger than
$\delta \eta_{\rm dec}$ and if, at the same time, $\theta_{\xi}
\gg \sigma$, where $\sigma$ is the detector angular resolution and
$\theta_{\xi} = \xi/(\eta_{0} -\eta_{\rm dec})$ represents the
angular dimension of a correlation domain on the surface of the
last scatter. Both effects of $\delta\eta_{\rm dec}\neq 0$ and of
$\sigma\neq 0$ reduce the calculated value for $RR'$. In this
sense, Eq.~(\ref{RR0}) should be regarded as an upper limit for
the true $RR'(0)$ value.

In order to estimate the role of $\delta \eta_{\rm dec}$ in the
calculation of $RR'(0)$, one can approximate the behavior of
$g(\eta)$ around the decoupling with $g(\eta)= (1 / \sqrt{2\pi}
\delta_{\rm dec}) \exp{[-(\eta-\eta_{\rm dec})^2/2\delta\eta_{\rm
dec}^2]}$, which is a Gaussian peaked at $\eta_{\rm dec}$ with a
characteristic width equal to $\delta\eta_{\rm dec}$. Introducing
this function into Eq.~(\ref{RRprimo}) one can calculate
explicitly the ratio between the rotation measure obtained taking
into account the thickness of the last scattering surface,
$\langle RR'(0)\rangle_{\delta \eta_{\rm dec}}$,  and the
``ideal'' value, $RR'(0)$, given by Eq.~(\ref{RR0}). One obtains
\begin{equation} \label{RRthick}
\frac{\langle RR'(0)\rangle_{\delta \eta_{\rm dec}}}{RR(0)} =
\frac{1}{2\sqrt{\pi}\delta \eta_{\rm dec}}\int_{-\infty}^{\infty}
d\eta' \exp \! \left(-\frac{\eta'^{2}}{4 \delta \eta_{\rm dec}^2}
\right) \frac{C_{\parallel}(\eta')}{C_{\parallel}(0)} \: .
\end{equation}
One should note that the thickness of the last scattering surface
introduces an average (along a line) of the function
$C_{\parallel}(r)$. This is easily explained if one considers the
physical meaning of this function. It describes (see
Eq.~(\ref{BBsymm})) the variations of the correlators $\langle
B_{i}(0) B_{i}({\bl r})\rangle$ of the magnetic field components
in a fixed direction (which in our case is the photon propagation
direction), when ${\bl r}$ moves parallel to this direction ({\it
i.e.} across the last scattering surface). For the specific case
$n_{S}=2$, we have
\begin{equation}
\frac{C_{\parallel}(\eta')}{C_{\parallel}(0)} = e^{-\frac{K^2
\eta'^2}{4}} ,
\end{equation}
which gives
\begin{equation} \label{RRded}
\frac{\langle RR'(0)\rangle_{\delta \eta_{\rm dec}}}{RR(0)} =
\left[ 1 + \frac{64 \pi}{9} \left(
 \frac{ \delta\eta_{\rm dec} }{\xi } \right)^{\!\!2} \,
\right] ^{-1/2} \!\!\! .
\end{equation}
For different values of the $n_{S}$ parameters, we obtain more
complicated expressions (which can be calculated analytically; see
the Appendix). For all the values of the $n_{S}$ parameter,
however, the ratio $\langle RR'(0)\rangle_{\delta \eta_{\rm
dec}}/RR'(0)$, for $\delta\eta_{\rm dec}\gg\xi$, always goes to
zero as $(\delta\eta_{\rm dec}/\xi)^{-1}$. In conclusion, since
$\Phi$ is proportional to $(RR')^{1/2}$ (see Eq.~(\ref{Phi}), we
find that the average rotation angle of the CMBR photons is
reduced with respect to Eq.~(\ref{RR0}) as $N^{-1/2}$, where $N =
\delta_{\rm dec}/\xi$ is the number of correlation domains in the
length $\delta\eta_{\rm dec}$.

\begin{figure}[htb]
\label{fig2} \plotone{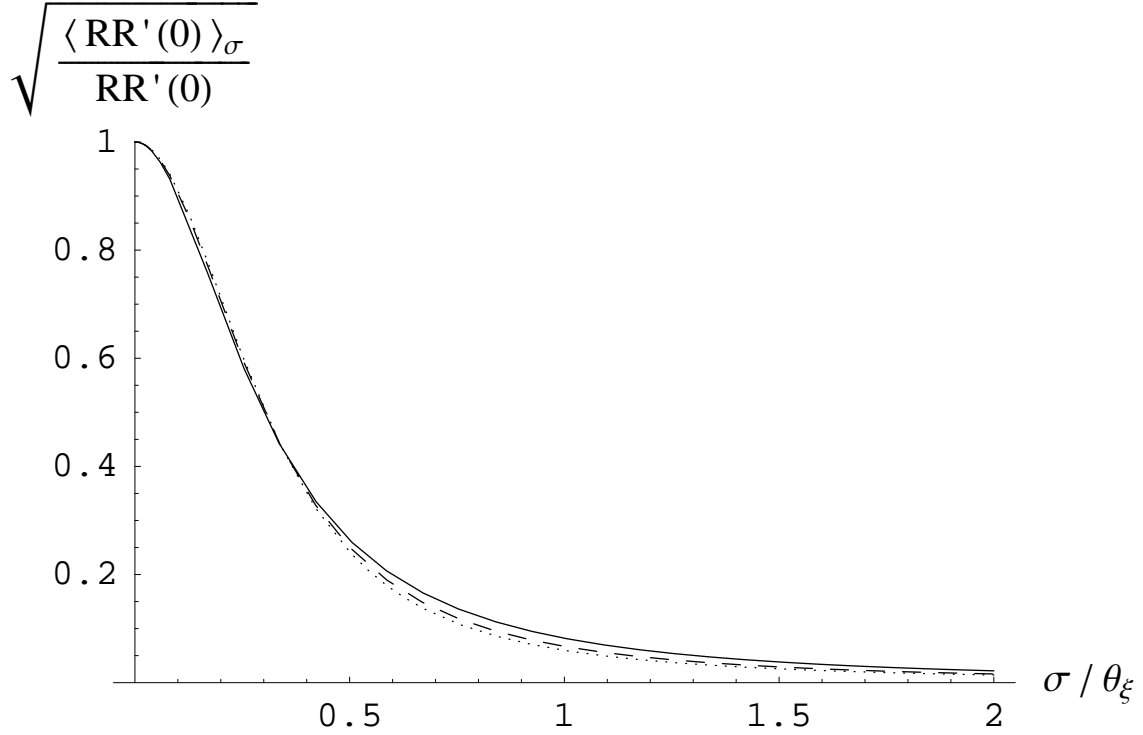} \caption{Effect of the detector
angular resolution $\sigma$ on the CMBR Faraday rotation measure.
The angle $\theta_{\xi}$ represents the angular dimension of a
magnetic field correlation domain on the surface of the last
scattering. The three lines correspond to the magnetic field
spectral index $n_S =2$ ({\it solid line}), $n_S = 4$ ({\it dashed
line}) and $n_S = 6$ ({\it dotted line}).}
\end{figure}

A similar exercise can be done to include the effect of finite
detector angular resolution. Modelling the detector angular
response with a Gaussian of angular width $\sigma$ (Kolb \& Turner
1990), one obtains
\begin{equation}
\frac{\langle RR'(0)\rangle_{\sigma}}{RR(0)} = \frac{1}{\sigma^2}
\int_{0}^{\infty} \! d\theta \: \theta \exp \!
\left(-\frac{\theta^{2}}{2 \sigma^2} \right)
\frac{C_{\perp}(\theta(\eta_0-\eta_{\rm dec}))}{C_{\perp}(0)} \: ,
\end{equation}
where $\langle RR'(0)\rangle_{\sigma}$ is the rotation correlator
obtained with an account of the detector angular resolution. We
have assumed here that $\sigma \ll 1$. In the case $n_{S}=2$, one
obtains
\begin{equation} \label{RRsigma}
\frac{\langle RR'(0)\rangle_{\sigma}}{RR(0)}= \left[ 1 + \frac{32
\pi}{9} \left( \frac{\sigma}{\theta_{\xi}} \right)^{\!\!2} \,
\right] ^{-2} \!\!\! ,
\end{equation}
where the angle $\theta_{\xi} = \xi/(\eta_{0} - \eta_{\rm dec})$
represents the angular dimension of a correlation domain on the
surface of the last scattering. For other $n_{S}$ values, we
obtain different expressions (see the Appendix) which, for $\sigma
\gg \theta_{\xi}$, always go to zero as
$(\theta_{\xi}/\sigma)^{-4}$. Thus, the rotation angle $\Phi$ is
reduced because of effect of the angular resolution as
$(\sigma/\theta_{\xi})^{-2}$. We remark that this behavior could
introduce severe limitations to the actual possibility of
observing the Faraday rotation of the CMBR polarization. In
particular, in Fig.2, we show $[\langle
RR'(0)\rangle_{\sigma}/RR'(0)]^{1/2}$ as a function of the ratio
$\sigma/\theta_{\xi}$ for different $n_{S}$ values. One sees that
even for small values of $\sigma/\theta_{\xi}$ the reduction of
the effect can be substantial.

We point out, finally, that in contrast to the previous case,
$\Phi$ is reduced as $N_{\sigma}^{-1}$ (and not as
$N_{\sigma}^{-1/2}$), where $N_{\sigma}=(\sigma/\theta_{\xi})^{2}$
is the number of correlation domains in a (two-dimensional) region
at the last scattering surface of angular dimension $\sigma$. This
is due to the peculiar behavior of the $C_{\perp}$ functions or,
in other words, to the fact that magnetic field components in
different regions of the last scattering surface can be
anticorrelated.

\section{Conclusions}

In this paper we have analyzed the Faraday rotation of the CMBR
polarization induced by a primordial stochastic magnetic field. In
particular, we have calculated the correlation between the Faraday
rotation measures, $RR'(\theta)$, as a function of the separation
angle between observation directions, $\theta$, for different
assumptions about the magnetic field spectral index and
correlation length. Here we summarize here our main results.

{\it i}) The helical part of the magnetic field does not
contribute to $RR'(\theta)$ at any angle $\theta$. This means that
Faraday RM maps can provide information only on the symmetric part
of the magnetic field.

{\it ii}) In the approach described in Sec.~3, neglecting the last
scattering surface thickness and the detector angular resolution,
the Faraday RM maps $RR'(\theta)$ can be calculated analytically.
We have provided an analytic expression for $RR'(\theta)$ for
arbitrary values of the magnetic field spectral index and
correlation length.

{\it iii}) The last scattering surface scale thickness $\delta
\eta_{\rm dec}$ reduces the Faraday rotation angle with respect to
the simple estimate given by Eq.~(\ref{Phi}). In the limit
$\delta\eta_{\rm dec}\gg\xi$, where $\xi$ is the magnetic field
correlation length, the rotation angle is reduced as $(\delta
\eta_{\rm dec}/\xi)^{-1/2}$ (see Eq.~(\ref{RRded})).

{\it iv}) The detector angular resolution $\sigma$  could
drastically reduce the possibility of observing the Faraday
rotation. In the limit $\sigma\gg\theta_{\xi}$, where
$\theta_{\xi}$ is the angular dimension of a magnetic field
correlation domain, the rotation angle is reduced with respect to
Eq.~(\ref{Phi}) as $(\sigma/\theta_{\xi})^{-2}$ (see
Eq.~(\ref{RRsigma})). As shown in Fig.~2, the reduction can be
substantial even for small values of $\sigma/\theta_{\xi}$.

\acknowledgements We would like to thank D. Comelli for helpful
discussions.

\clearpage

\appendix
\section{The $C_\perp (r)$ and
$C_\parallel (r)$ correlators}

In this appendix, we calculate the correlators $C_\perp (r)$ and
$C_\parallel (r)$ defined by Eqs.~(\ref{c-perp}) and
(\ref{c-parallel}), for the functional form of the magnetic power
spectrum given by Eq.~(\ref{spectrum}). To this end, we introduce
the function
\begin{equation}
L_{n_S}(\alpha) = \frac{\int_{0}^{\infty} dk k^{n_s + 2} e^{-
\alpha k^2/K^2} \left[ \, j_0(kr)- j_2(kr) / 2 \, \right]}{
\int_{0}^{\infty} dk k^{n_s + 2} e^{-k^2/K^2}} \: ,
\end{equation}
related to $C_\perp (r)$ by
\begin{equation} \label{L0}
L_{n_S}(0) = \frac{C_\perp (r)}{C_\perp (0)} \, .
\end{equation}
It is straightforward to obtain the recursion formula for
$L_{n_S}(\alpha)$
\begin{equation} \label{recursion}
L_{n_S}(\alpha) =  A_m \frac{\partial^m}{\partial \alpha^m}
L_{2}(\alpha) \, ,
\end{equation}
where the ``generating function'' $L_{2}(\alpha)$ is given by
\begin{equation} \label{generating}
L_2(\alpha) = \frac{e^{-\kappa^2 r^2/\alpha}}{\alpha^{5/2}} \left(
1 - \frac{\kappa^2 r^2}{\alpha} \right) \! ,
\end{equation}
and
\begin{equation} \label{constants}
m = \frac{n_S}{2}-1 , \;\;\;\; \kappa = \frac{K}{2} , \;\;\;\; A_n
= \frac{3\,(-1)^n \, 2^n}{(2n+3)!!}
\end{equation}
(we remember that $n_S \geq 2$ is an even natural number).
\\
Taking into account Eqs.~(\ref{recursion})-(\ref{generating}) we
cast Eq.~(\ref{L0}) in the form
\begin{equation} \label{cperp}
\frac{C_\perp (r)}{C_\perp (0)} = e^{-\kappa^2 r^2}
P_{\frac{n_s}{2}-1}({\kappa^2 r^2}) \, ,
\end{equation}
where $P_n(x)$ turn out to be polynomials of $(n+1)$st degree
defined by
\begin{equation}
\label{Polinomi di Campanelli-Giannotti1} P_n(x) = A_n \: e^{x}
\frac{\partial^n}{\partial \alpha^n} \,
\frac{e^{-x/\alpha}}{\alpha^{5/2}} \left( 1- \frac{x}{\alpha}
\right) |_{\alpha = 1} \: , \;\;\; n = 0,1,2,...
\end{equation}
The first three $P_n$ polynomials are
\begin{eqnarray}
P_0(x) \!\!& = & \!\! 1 - x , \nonumber \\
P_1(x) \!\!& = & \!\! 1 - \frac{9}{5} \, x + \frac{2}{5} \, x^2 , \\
P_2(x) \!\!& = & \!\! 1 - \frac{13}{5} \, x + \frac{8}{7} \, x^2 -
\frac{4}{35} \, x^3 . \nonumber
\end{eqnarray}
Following the same procedure, we find for $C_\parallel (r)$ the
expression
\begin{equation} \label{cpar}
\frac{C_\parallel (r)}{C_\parallel (0)} = e^{-\kappa^2 r^2}
Q_{\frac{n_s}{2}-1}({\kappa^2 r^2}) \, ,
\end{equation}
where $Q_n(x)$ are polynomials of $n$th degree given by
\begin{equation}
\label{Polinomi di Campanelli-Giannotti2} Q_n(x) = A_n \: e^{x}
\frac{\partial^n}{\partial \alpha^n} \,
\frac{e^{-x/\alpha}}{\alpha^{5/2}} |_{\alpha = 1} \: , \;\;\; n =
0,1,2,... ,
\end{equation}
where $\kappa$ and $A_n$ are the same as in Eq.~(\ref{constants}).
We give the expressions for the first three $Q_n$ polynomials:
\begin{eqnarray}
Q_0(x) \!\!& = & \!\! 1 , \nonumber \\
Q_1(x) \!\!& = & \!\! 1 - \frac{2}{5} \, x  , \\
Q_2(x) \!\!& = & \!\! 1 - \frac{4}{5} \, x + \frac{4}{35} \, x^2 .
\nonumber
\end{eqnarray}
Now, having the expressions for $C_\perp (r)$ and $C_\parallel
(r)$, we can easily calculate the correlators defined by
Eqs.~(\ref{RRthick}) and (\ref{RRsigma}). Taking into account
Eqs.~(\ref{cpar})-(\ref{Polinomi di Campanelli-Giannotti2}), we
cast Eq.~(\ref{RRthick}) in the form
\begin{equation}
\frac{\langle RR'(0)\rangle_{\delta \eta_{\rm dec}}}{RR'(0)} = A_m
\frac{\partial^m}{\partial \alpha^m} \, \frac{1}{\alpha^{2}
\sqrt{\alpha + (K \delta \eta_{\rm dec})^2}} \: |_{\alpha = 1} \:
,
\end{equation}
while, inserting Eqs.~(\ref{cperp})-(\ref{Polinomi di
Campanelli-Giannotti1}) into Eq.~(\ref{RRsigma}), we get
\begin{equation}
\frac{\langle RR'(0)\rangle_{\sigma}}{RR'(0)} = A_m
\frac{\partial^m}{\partial \alpha^m} \, \frac{1}{\sqrt{\alpha}
\left( \alpha + 2\chi^2 \sigma^2 \right)^2} \: |_{\alpha = 1} \: ,
\end{equation}
where we remember that $\chi = K(\eta_0 - \eta_{\rm dec})/2$, and
$m$ is given by Eq.~(\ref{constants}).
\\
Finally, we give the asymptotic expressions
\begin{equation}
\delta \eta_{\rm dec} \gg \xi : \;\;\;\; \frac{\langle
RR'(0)\rangle_{\delta \eta_{\rm dec}}}{RR'(0)} = \frac{3}{8
\sqrt{\pi}} \, \left( \frac{\delta \eta_{\rm dec}}{\xi}
\right)^{\!\! -1} \! ,
\end{equation}
\begin{equation}
\sigma \gg \theta_\xi : \;\;\;\; \frac{\langle
RR'(0)\rangle_{\sigma}}{RR'(0)} = C_m \! \left(
\frac{\sigma}{\theta_\xi}\right)^{\!\! -4} \! ,
\end{equation}
where we remember that $\xi$ is the correlation length (see
Eq.~(\ref{xi})), the coefficients $C_m$ are
\begin{equation}
C_m = \frac{3}{(2 \pi)^4} \, \frac{\Gamma(m + 1/2) \left[ \,
\Gamma(m + 5/2) \, \right]^{\,3}}{\left[ \, \Gamma(m+2) \,
\right]^{\,4}} \, ,
\end{equation}
and $\theta_\xi = \xi/(\eta_0 - \eta_{\rm dec})$ is the angular
dimension of a correlation domain on the surface of the last
scatter.

\end{document}